\pdfoutput=1
\documentclass[manuscript]{aastex}
\usepackage{url}
\usepackage{bm}
\usepackage{amsmath}
\usepackage{appendix}

\RequirePackage{color}

\begin{document}

\title{Path integral approach for merger rates of dark matter haloes}
\shortauthors{N. Hiotelis}

\author{N. Hiotelis \altaffilmark{1}}
\affil{1st  Experimental Lyceum of Athens, Ipitou 15, Plaka, 10557,
Athens,  Greece}
\email{hiotelis@ipta.demokritos.gr}
\altaffiltext{1}{Lysimahias 66, Neos Kosmos, Athens, 11744 Greece,
e-mail:hiotelis@ipta.demokritos.gr }

\begin{abstract}
   We use path integrals in order to estimate merger rates of dark matter haloes using the Extended Press-Schechter approximation (EPS) for the Spherical Collapse (SC) and the Ellipsoidal Collapse (EC) models.\\
   Merger rates have been calculated for masses in the range $10^{10}M_{\odot}\mathrm{h}^{-1}$ to $10^{14}M_{\odot}\mathrm{h}^{-1}$ and for redshifts  $z$ in the range $0$ to $3$. A detailed comparison between these models is presented. Path approach gives a  better agreement with the exact solutions for constrained distributions than the approach of \cite{shto02}.  Although this improvement seems not to be very large, our results show that the path approach is a step to the right direction. Differences between the two widely used barriers, spherical and ellipsoidal, depend crucially on the mass of the descendant halo. These differences become larger for decreasing mass of the descendant halo.\\
     The use of additional terms in the expansion used in the path approach, other improvements  as well as  detailed comparisons with the predictions of N-body simulations, that could improve our understanding about the important issue of structure formation, are under study.
\end{abstract}


\section{Introduction}
The development of analytical or semi-numerical methods for the problem of structure
formation in the universe helps to improve our understanding of important physical processes. A class
of such methods is based on the ideas of
\citet{prsc74} and on their extensions. These extensions are called Extended Press-Schechter Methods (EPS),   and they are presented in the pioneered works of  \citet{pea}, \citet{boet91} and \citet{laco93}.\\
In this section we summarize useful relations about density perturbations, filters,  stochastic processes and path integrals that are used in the following calculations.\\
The overdensity at a given point $\textbf{r}$ of the initial Universe is given by the relation:
\begin{equation}
\delta(\textbf{r})\equiv \frac{\rho(\textbf{r})-\rho_b(\textbf{r})}{\rho_b(\textbf{r})}
\end{equation}
 In the above relation, $\rho(\textbf{r})$ is the density at point $\textbf{r}$ of the initial Universe.
Index $b$ denotes the density of the background model of the Universe.
The smoothed density perturbation at $\textbf{r}$ is defined by the relation:
\begin{equation}
\delta(\textbf{r};R)=\int\delta(\textbf{x})W(\textbf{r}-\textbf{x};R)\mathrm{d}^3\textbf{x}
\end{equation}
where $W$ is a filter with characteristic radius $R$.
Using the convolution theorem to transform both sides, we have:
\begin{equation}
\hat{\delta}(\textbf{k};R)=\hat{\delta}(\textbf{k})\hat{W}(\textbf{k};R)
\end{equation}
In our calculations we use the sharp in $\textbf{k}$ space filter given by:
\begin{eqnarray}
W_{KS}(r;R)=\frac{1}{6{\pi}^2R^3}\frac{3( \sin x-x\cos x)}{x^3},~~x\equiv r/R \nonumber\\
\hat{W}_{KS}(k;R)
=H(1-kR)
\end{eqnarray}
where $x\equiv r/R$ and $H$ is the Heaviside step function defined by:
\begin{equation}
H(x)=\left\{
\begin{array}{l l}
0,~x<0\\
\frac{1}{2},~x=0\\
1,~x>0
\end{array}\right.
\end{equation}
 For spherically symmetric kernels, such  the ones we examine here,  the variance of the overdensity at scale $R$ is given by :
 \begin{equation}
 S(k_f)\equiv \sigma^2(k_f)=\int_{-\infty}^{+\infty}\Delta^2(k)|\hat{W}(k;k_f)|^2\mathrm{d}\ln k
 \end{equation}
 where $k_f\equiv 1/R$ and ${\Delta}^2=\frac{k^3}{(2\pi)^3}P_s(k)$.  $P_s(k)$ is the power spectrum .
 For the k-sharp filter the evolution of smoothed $\delta$  as a function of $S$  is governed by a Langevin  equation:
 \begin{equation}
 \frac{\mathrm{d}\delta}{\mathrm{d}S}=n(S)
 \end{equation}
 \begin{equation}
 <n(S)n(S')>= \delta_D(S-S')
 \end{equation}
 (\cite{lan}, \cite{cof}), where $\delta_D$ is the Dirac delta function,  that describes the following Markovian stochastic process: Trajectories start from the same position $(S_0, \delta_0)$ at the $(S, \delta)$ plane and evolve according to Eq.7 as $S$ decreases. $S$ is completely analogous to time $t$ in ordinary problems involving stochastic processes. The probability  $P(S,\delta/S_0,\delta_0)\mathrm{d}\delta$,  a trajectory that starts from the position $(S_0,\delta_0)$ passes at $S$ from a value of $\delta$ in the interval  $[\delta,\delta+\mathrm{d}\delta]$ satisfies a Fokker - Planck equation (Coffey et al. 2005), that leads to the diffusion equation:
 \begin{equation}
 \frac{\partial P(S,\delta/S_0,\delta_0)}{\partial S}=\frac{1}{2}\frac{\partial ^2}{\partial \delta^2}\left[P(S,\delta / S_0,\delta_0)\right]
 \end{equation}
 with solution:
    \begin{equation}
    P(S,\delta/S_0,\delta_0)=\frac{1}{\sqrt{2\pi(S-S_0)}}\exp\left[-\frac{(\delta-\delta_0)^2}{2(S-S_0)}\right]
    \end{equation}
    In Sect.2 we give  the basic equations from the path integral approach method. In Sect.3 first crossing distributions and structure formation are discussed. In Sect. 4 analytical relations for merger rates and mean merger rates are presented. Results for different models as well as comparisons are presented. In Sect. 5 we summarize the method, the results and we give a short discussion.
    \section{Path integral approach: Basic Equations}
     Path integrals are power tools for the study of various fields of theoretical physics, and they have been studied extensively in the literature, (\citet{wi} \citet{fey},\citet{gro}, \citet{cha}). They  are given by the sum over all possible paths, satisfying some boundary conditions. Applications appeared for the problem of structure formation in the Universe from \citet{boet91}, \citet{magg} and \citet{sim}. We follow the formalism of the above authors.\\
    According to the above described picture for the formation of structures, we consider an ensemble of trajectories all starting from the point $(S_0,\delta_0)$
    and we follow their evolution for the ``time" interval $[S_0,S]$. The interval is discretized in steps $\Delta S$ such as $S_k=S_0+k\epsilon$ where $k=1,2...n$ and $S_n=S$. A trajectory is defined by the points $(S_0,\delta_0),(S_1,\delta_1)...(S_n,\delta_n)$. The probability density the variables $\delta(S_1),\delta(S_2)...\delta(S_n)$ take the values $\delta_1,\delta_2...\delta_n$ respectively is:
    \begin{equation}
    W(S_0,\delta_0;S_n,\delta_1,\delta_2...\delta_n)\equiv\langle\delta_D(\delta(S_1)-\delta_1)...
    \delta_D(\delta(S_n)-\delta_n)\rangle
    \end{equation}
    Using the integral representation of  the Dirac delta function :
    \begin{equation}
    \delta_D(x)=\int_{-\infty}^{+\infty}\frac{\mathrm{d}\lambda}{2\pi}e^{-i\lambda x}
    \end{equation}
     and the fact that  each one of the variables $\delta(S_1),\delta(S_2)...\delta(S_n)$ follows a central Gaussian distribution, the probability $\Pi_{\epsilon}$ of arriving at $(S_n, \delta_n)$ starting from $(S_0,\delta_0)$ using discrete trajectories of step $\epsilon$ that never exceeded a barrier $B$ that depends on $S$ is:
      \begin{equation}
    {\Pi}_{\epsilon}(S_0,\delta_0;S_n,\delta_n)=\int_{-\infty}^{B_n}\mathrm{d}\delta'_1
    \int_{-\infty}^{B_n}\mathrm{d}\delta'_2...
    \int_{-\infty}^{B_n}\mathrm{d}\delta'_{n-1}W(S_0,\delta_0;S_n,\delta'_1,\delta'_2...\delta'_n)
    \end{equation}
    where,
    \begin{equation}
    W(S_0,\delta_0;S_n,\delta'_1,\delta'_2...\delta'_n)=
     \int_{-\infty}^{+\infty}\frac{\mathrm{d}\lambda_1}{2\pi}\int_{-\infty}^{+\infty}\frac{\mathrm{d}\lambda_2}{2\pi}...
    \int_{-\infty}^{+\infty}\frac{\mathrm{d}\lambda_n}{2\pi}e^{z'}
     \end{equation}
     and
     \begin{eqnarray}
      z'= i\sum_{k=1}^{n}\lambda_k(B_k-B_n)+ i\sum_{k=1}^{n}\lambda_k\delta'_k-\frac{1}{2}\sum_{k=1}^{n}\sum_{j=1}^{n}\lambda_k\lambda_jS_{kj}
     \end{eqnarray}
     Expanding  in a Taylor series the quantity $e^{i\sum_{k=1}^{n}\lambda_k(B_k-B_n)}$ as well as the difference $B_k-B_n$ around $S_n$ (see \cite{ls} and the discussion therein),  we have $W=W^{(0)}+W^{(1)}+W^{(2)}+...$, where
     \begin{eqnarray}
     W^{(0)}=
     \frac{1}{(2\pi)^n}\int_{-\infty}^{+\infty}{\cal{D}\bm{\lambda}}
     \exp[i\sum_{k=1}^{n}\lambda_k\delta_k-
     \frac{1}{2}\sum_{k=1}^n\sum_{j=1}^n\lambda_k\lambda_jS_{ij}]
     \end{eqnarray}
     \begin{eqnarray}
     W^{(1)}=
     \frac{i}{(2\pi)^n}\sum_{k=1}^nC_k\int_{-\infty}^{+\infty}{\cal{D}\bm{\lambda}}\lambda_k
     \exp[i\sum_{k=1}^{n}\lambda_k\delta_k-
     \frac{1}{2}\sum_{k=1}^n\sum_{j=1}^n\lambda_k\lambda_jS_{ij}]
     \end{eqnarray}
     \begin{eqnarray}
     W^{(2)}=
     -\frac{1}{2(2\pi)^n}\sum_{k=1}^n\sum_{j=1}^nD_{kj}\int_{-\infty}^{+\infty}{\cal{D}\bm{\lambda}}\lambda_k\lambda_j
     \exp[i\sum_{k=1}^{n}\lambda_k\delta_k-
     \frac{1}{2}\sum_{k=1}^n\sum_{j=1}^n\lambda_k\lambda_jS_{ij}]
     \end{eqnarray}
      where we set:
     \begin{equation}
     \int_{-\infty}^{+\infty}{\cal{D}\bm{\lambda}}\equiv\int_{-\infty}^{+\infty}\mathrm{d}\lambda_1..
     \int_{-\infty}^{+\infty}\mathrm{d}\lambda_n
     \end{equation}
     \begin{equation}
     C_k=\sum_{p=1}^{+\infty}\frac{B_n^{(p)}}{p!}(S_k-S_n)^p
     \end{equation}
     and:
     \begin{equation}
     D_{kj}=\sum_{p=1}^{+\infty}\sum_{q=1}^{+\infty}\frac{B_n^{(p)}B_n^{(q)}}{p!q!}(S_k-S_n)^p(S_j-S_n)^q
     \end{equation}
     The symbol $B_n^{(p)}$ denotes $\frac{\mathrm{d}^pB(S)}{\mathrm{d}S^p}/_{S=S_n}$.\\
         It is straightforward to show that $W^{(1)}$ and $W^{(2)}$ are connected to $W^{(0)}$,by:
      \begin{equation}
      W^{(1)}=\sum_{k=1}^{n}{\partial_kW^{(0)}},~~~~W^{(2)}=-\frac{1}{2}\sum_{k=1}^n\sum_{j=1}^nD_{kj}\partial_k\partial_jW^{(0)}
      \end{equation}
      where $\partial_k \equiv\frac{\partial }{\partial \delta_k}$ and $\partial_{k_1}\partial_{k_2}..\partial_{k_l}\equiv\frac{{\partial}^l}{\partial \delta_{k_1}\delta_{k_2}...\delta_{k_l}}$.
      The quantity  $W^{(0)}$ is known as Wiener measure and is given by:
      \begin{equation}
      W^{(0)}(S_0,\delta_0=0;S_n,\delta_1,\delta_2...\delta_n)=\frac{1}{(2\pi\epsilon)^{n/2}}\exp[-\frac{1}{2\epsilon}
      \sum_{k=1}^{n}(\delta_k-\delta_{k-1})^2]
      \end{equation}
                       We note that the integral $\int_{-\infty}^{B(S_n)}\Pi_{\epsilon=0}(S_0,\delta_0;S_n,\delta_n)\mathrm{d}\delta_n$ equals to the number density of trajectories that have never crossed the barrier. This number density is a decreasing value of ``time" $S$, since the number of trajectories that pass the boundary increases with increasing $S$. However, the rate of change of the above integral shows the number of trajectories that cross - for first time- the barrier at  $S$. This is the first crossing distribution $\cal{F}$ that satisfies the Eq.
      \begin{equation}
      {\cal{F}}(S_n)=-\frac{\partial}{\partial S_n}\int_{-\infty}^{B(S_n)}\Pi_{\epsilon=0}(S_0,\delta_0;S_n,\delta_n)
      \mathrm{d}\delta_n
      \end{equation}
       Assuming that every component  (16), (17) and (18) satisfies a diffusion equation, we can write:

        \begin{equation}
        {\cal{F}}^{(j)}(S_n)=-\frac{1}{2}\frac{\partial}{\partial\delta_n}\Pi^{(j)}_{\epsilon=0}(S_0,\delta_0;S_n,\delta_n)|{_{\delta_n=B(S_n)}}, j=0,1
        \end{equation}
        Finally  we have:
        \begin{eqnarray}
        {\cal{F}}^{(0)}(S_n)=\frac{B_n-\delta_0}{\sqrt{2\pi}(S_n-S_0)^{3/2}}\exp\left[-\frac{[B_n-\delta_0]^2}{2(S_n-S_0)}\right]\nonumber\\
        {\cal{F}}^{(1)}(S_n)=\frac{1}{2\pi}(B_n-\delta_0)\sum_{p=1}^{+\infty}\frac{(-1)^p}{p!}B^{(p)}_n\int_{S_0}^{S_n}\frac{(S_n-S_l)^{p-3/2}}{(S_l-S_0)^{3/2}}\exp[-\frac{(B_n-\delta_0)^2}{2(S_l-S_0)}]\mathrm{d}S_l
        \end{eqnarray}
        that in terms of the confluent hypergeometric function $U$ (\cite {abra}) can be written:
         \begin{eqnarray}
         {\cal{F}}^{(1)}(S_n)=\frac{(B_n-\delta_0)}{2\pi}e^{-\psi}
         \sum_{p=1}^{+\infty}(S_n-S_0)^{p-2}\frac{(-1)^p}{p!}B^{(p)}_n
         \Gamma\left(p-\frac{1}{2}\right)U\left(p-\frac{1}{2},\frac{3}{2},\psi\right)=\nonumber\\
         \frac{e^{-\psi}}{\pi \sqrt{2}} \sum_{p=1}^{+\infty}
         (S_n-S_0)^{p-\frac{3}{2}}\frac{(-1)^p}{p!}B^{(p)}_n\Gamma\left(p-\frac{1}{2}\right)
         U\left(p-1,\frac{1}{2},\psi\right)
         \end{eqnarray}
         where $\psi\equiv\frac{(B_n-\delta_0)^2}{2(S_n-S_0)}$ and $\Gamma$ is the complete gamma function.\\
         Thus ${\cal{F}}(S_n)\equiv{\cal{F}}^{(0)}(S_n)+{\cal{F}}^{(1)}(S_n)$ is written in the form:
         \begin{equation}
         {\cal{F}}_{Path}(S,B(S)/S_0,\delta_0)=\frac{e^{-\psi}}{\sqrt{(2\pi)}(S-S_0)^{3/2}}G(S,S_0,B,\delta_0)
         \end{equation}
         where the quantity $G$ is defined by the relation:
         \begin{equation}
         G(S,S_0,B,\delta_0)\equiv
         B(S)-\delta_0+\sum_{p=1}^{+\infty}\frac{(S_0-S)^p}{p!\sqrt{\pi}}
         B^{(p)}(S)\Gamma(p-\frac{1}{2})U\left(p-1,\frac{1}{2},\psi\right)
         \end{equation}
         In the above relation we have set $S=S_n$, $B(S)=B_n $, $U(0,a,b)=1$ and we rewrote the arguments in a different form, useful in what follows.\\
         It is also useful for our purposes to consider a barrier that is a function of redshift $z$ too, $B\equiv B(S,z)$. Additionally, we also assume that the value $\delta_0$ satisfies the relation $\delta_0\equiv B(S_0,z_0)$. However, the constraint $(S,B(S,z)/ S_0,\delta_0)$ is equivalent
          to $(S,B(S,z)/ S_0,B(S_0,z_0))$ or without any confusion $(S,z/ S_0,z_0)$. Thus, given that a trajectory crosses, for first time,  the barrier $B(S_0,z_0)$ at redshift $z_0$, ${\cal{F}}(S,z,/S_0,z_0)$ gives the probability this trajectory to cross, for first time,  the barrier $B(S,z)$ at $z$. We also denote ${\cal{F}}(S,z)\equiv{\cal{F}}(S,z/S_0=0,z=z_{in})$, where $z_{in}$ satisfies $B(0,z_{in})=0$.\\
           Obviously, in the above relations $B(S)$ has to be replaced by  $B(S,z)$ and $B^{(p)}(S)$ by $B^{(p)}(S,z)\equiv \frac{\partial^p}{\partial S^p}B(S,z)$. Thus, it is more convenient to write:
         \begin{equation}
         G(S,S_0,z,z_0)\equiv
         B(S,z)-B(S_0,z_0)+\sum_{p=1}^{+\infty}\frac{(S_0-S)^p}{p!\sqrt{\pi}}
         B^{(p)}(S,z)\Gamma(p-\frac{1}{2})U\left(p-1,\frac{1}{2},\psi\right)
         \end{equation}
         The use of the above formula in practice, requires taking account a finite number of terms. Due to the behavior of the confluent hypergeometric function, it is not obvious that the leading order term dominates the sum. In our calculations, we tried an increasing number of terms $N_T$ (thus the sum above extends from $p=1$ to $p=N_T$). We found that for $N_T \geq 12$ the results are the same. So, the value of $N_T =12$ is sufficient for our purposes. We note that we also checked the results for very large number of $N_T$ as $100$ and no differences were found.

         \section{First crossing distributions and structure formation}

          First crossing distributions are connected to structure formation (\cite{boet91}, \cite{laco93}). We consider the variable $\hat{M}$ that is the relative excess of mass at scale $R$. This is written in the form:
         \begin{equation}
         \hat{M}(R)\equiv\frac{M(R)-M_b(R)}{M_b(R)}=\frac{3}{R^3}\int_0^R\delta(r)r^2\mathrm{d}r
         \end{equation}
         where $M(R)$ is the mass contained in a sphere of radius $R$ of the Universe and $M_b(R)=\frac{4}{3}\pi \rho_bR^3$ is the mass contained in a sphere of radius $R$ of the unperturbed model of the Universe that has a constant density $\rho_b$.\\
         It is easy to check that in the case $\delta$ follows a central Gaussian, then $\hat{M}$ follows a central Gaussian  with the same variance. Thus, it is reasonable to assume that the relation between $S$ and $R$ (see Eq. 6) can be transformed into a relation between $S$ and $M$. This assumption should have a complete physical meaning if the volume associated with the filter was that of a sphere of radius R. The volume associated with a filter $F$ is given by:
          \begin{equation}
         V_{F}=\frac{4}{3}\pi R^3\int_0^{+\infty} 4\pi W_{F}(r;R)r^2\mathrm{d}r
          \end{equation}
          The Gaussian filter has an infinite extent and so it is difficult to understand the physical connection between $R$ and $M$ and for the k-sharp filter the above integral does not exist at all and so the volume is not even well defined. However, since the k-sharp filter is so convenient for the analytical approximation of the problem studied here, we followed the usual procedure that is to assume a mass $M=\frac{4}{3}\pi\rho_b R^3$ associated with the k-sharp filter. We note that in \cite{laco93} a volume $V=6\pi^2R^3$ is quoted for the k-space filter, a result that is used without much justification, see the details in \cite{magg}.\\
       The second point is to connect the first crossing distribution of trajectories with the number density of haloes. This is done by using the following argument: The probability a mass element at redshift $z$ belongs to a halo of mass in the range $M, M+\mathrm{d}M$ denoted by $f(M,z)\mathrm{d}M$ equals to the probability a trajectory crosses, for first time, the barrier $B(S,z)$ between $S,S+\mathrm{d}S$ denoted by ${\cal{F}}(S,z)\mid\mathrm{d}S\mid$. Variables $S$ and $M$ are connected by the relation $S=\sigma^2(M)$. The described equation can be written in the form:
          \begin{equation}
           f(M,z)\mathrm{d}M={\cal{F}}(S,z)\mid\frac{\mathrm{d}S}{\mathrm{d}M}\mid\mathrm{d}M
          \end{equation}
          and since we assume that all trajectories start from the point $(S_0=0,\delta_0=0)$ this is an unconstrained probability.
          For the constrained case, we write:
           \begin{equation}
           f(M,z/M_0,z_0)\mathrm{d}M={\cal{F}}(S,z/S_0,z_0)\mid\frac{\mathrm{d}S}{\mathrm{d}M}\mid\mathrm{d}M
          \end{equation}
          A form of the barrier that results to a mass function that is in good agreement with the results of N-body simulations is $B_{EC}(S,z) $ given by:
  \begin{equation}
  \frac{B_{EC}(S,z)}{\sqrt{\alpha}B_{SC}(z)}=1+\frac{\beta}{[\alpha B_{SC}^2(z)]^{\gamma}}S^{\gamma}
  \end{equation}
  In the above Eq. $\alpha$, $\beta$ and $\gamma$ are constants.
  The above barrier represents an ellipsoidal  collapse model (EC),
  \citet{smt}, \citet{shto02} (ST02 hereafter). The barrier  depends on the mass ($S=S(M)$) and it is called a ``moving barrier".
  The values of the  parameters are $\alpha=0.707$,~$\beta=0.485$,~ $\gamma=0.615$ and are adopted
    either from the dynamics of ellipsoidal collapse or from
  fits to the results of N-body simulations.  In the above relation  $B_{sc}(z)=1.686/D(z)$, where $D(z)$ is the growth factor derived by the linear theory, normalized to unity at the present epoch. $B_{sc}(z)$ is the barrier for the spherical collapse model that results from the above relation for  $\alpha=1$ and $\beta=0$. In the plane $(S,\delta)$ the line $\delta_c=B_{SC}(z)$  is parallel to the $S$ axis. The physical picture is that in an Einstein-de Sitter Universe a spherical region collapses at $z$ if the linear extrapolation of its initial value $\delta_{in}$ up to the present epoch equals to $B_{SC}(z)$ (see for example \citet{peeb80}).\\
  The above relation can be written in a more convenient form:
  \begin{equation}
  B_{EC}(S,z)=p_c(z)+q_c(z)S^{\gamma}
  \end{equation}
  where $p_c(z)= w_1B_{SC}(z),q_c(z)=w_2B^{-w_3}_{SC}(z)$ with $w_1=\sqrt{\alpha},w_2=\beta{\alpha}^{0.5-\gamma},w_3=2\gamma-1$.
  For the spherical collapse (SC) model, $\cal{F}$ can be calculated analytically. The analytical solution for the SC model is given by:
  \begin{equation}
 {\cal{F}}_{SC}(S,z)=\frac{p_c(z)}{\sqrt{2\pi S^3}}\exp\left[-\frac{p_c^2(z)}{2S}\right]
  \end{equation}
  \citet{shto02} have shown that a good approximation of $ {\cal{F}}$ for the EC model is given by the relation:
   \begin{equation}
 {\cal{F}}_{EC-ST}(S,z)=\frac{1}{\sqrt{2\pi S^3}}\mid T(S,z)\mid\exp\left[-\frac{B^2_{EC}(S,z)}{2S}\right]
  \end{equation}
  where
  \begin{equation}
  T(S,z)=B_{EC}(S,z)+\sum_{k=1}^{5}\frac{(-S)^k}{k!}\frac{\partial^k}{\partial S^k}B_{EC}(S,z)
  \end{equation}
  We note that for the constrained case the above relations are written:
  \begin{equation}
 {\cal{F}}_{SC}(S,z/S_0,z_0)=\frac{B_{SC}(z)-\delta_0}{\sqrt{2\pi (S-S_0)^3}}\exp\left[-\frac{[B_{SC}(z)-\delta_0]}{2(S-S_0)}\right]
  \end{equation}
  and
  \begin{equation}
 {\cal{F}}_{EC-ST}(S,z/S_0,z_0)=\frac{1}{\sqrt{2\pi (S-S_0)^3}}\mid T(S,z/S_0,z_0)\mid\exp\left[-\frac{[B_{EC}(S,z)-\delta_0]^2}{2(S-S_0)}\right]
  \end{equation}
  with
  \begin{equation}
  T(S,z/S_0,z_0)=B_{EC}(S,z)-\delta_0+\sum_{k=1}^{5}\frac{(S_0-S)^k}{k!}\frac{\partial^k}{\partial S^k}B_{EC}(S,z)
  \end{equation}
  Comparing the expressions (28) and (41) describing the path approach and the ST02 approximation respectively, it is obvious that (41) results from (28) by just replacing $G$ with $\mid T\mid$. A comparison between the expressions of $G$ and $ T$, that are given by Eqs (29) and (42) respectively, shows that their leading terms, for $p=1$ and $k=1$ respectively, are equal. As regards the whole sums we can draw some conclusions in the case of small $\psi$. Using the asymptotic formula 13.5.10, $U(a,b,\psi)=\frac{\Gamma(1-b)}{\Gamma(1+a-b)}+{\cal{O}}(\mid\psi\mid^{1-b})$,  of \cite{abra} that holds for small $\psi$ and $0 < b <1$ , for $a=p-1$ and $b=1/2$ and substituting in (29) we see that $G$ is close to $T$.
  \subsection{Mass functions and N-body simulations}
       On the other hand, large numerical simulations give very important information about mass functions. Such simulations, starting from cosmological initial conditions, can find at different values of the redshift $z$, the number density of haloes of given mass $M$, denoted by $N(M,z)$, and the variance of mass  at scale $M$, denoted by $S(M)$. It is well known that these quantities are related by the equation:
     \begin{figure}
\includegraphics[width=16cm]{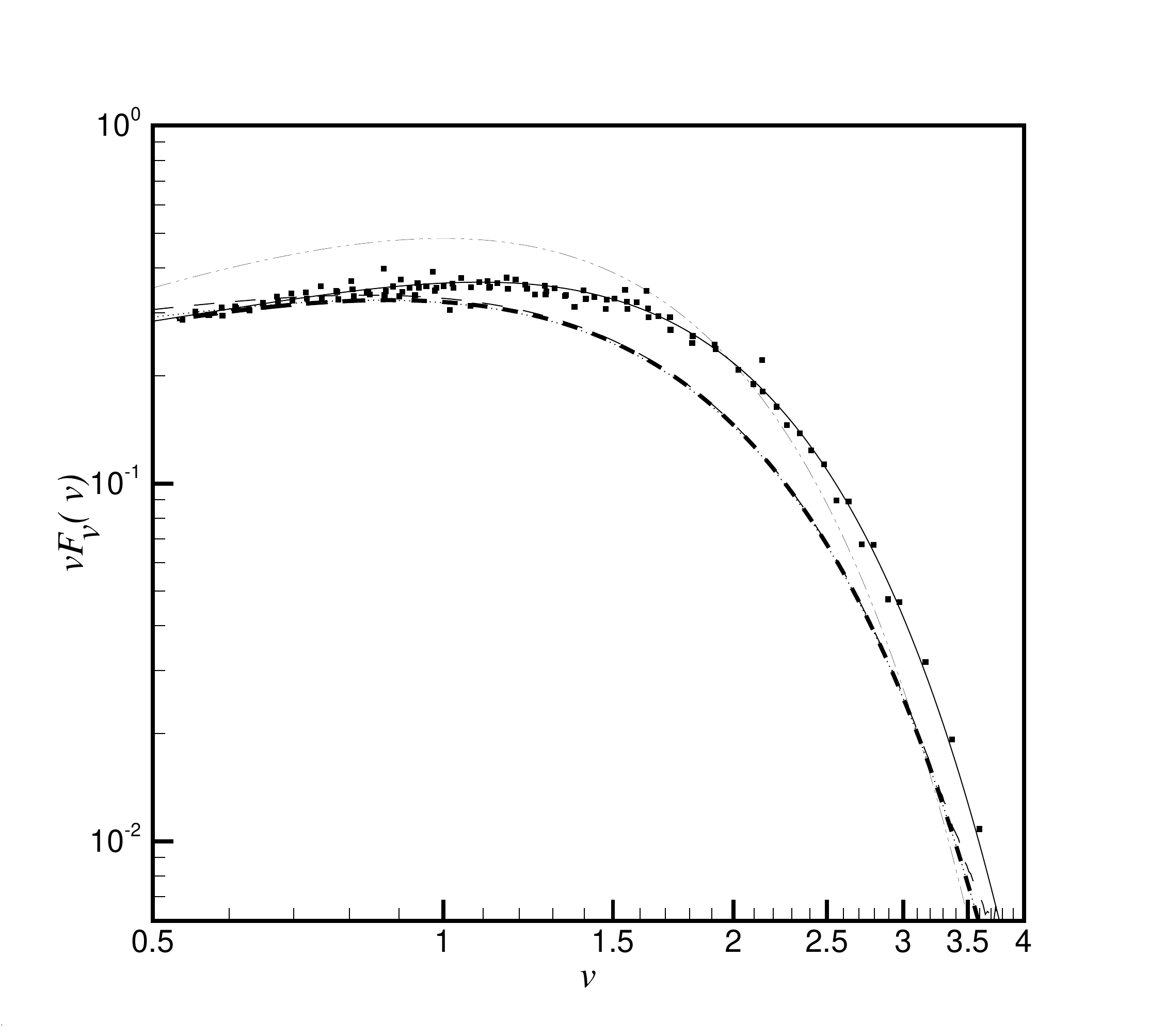}
\caption{ Multiplicity functions, $\nu{\cal{F}}_{\nu}(\nu)$ satisfying (47). Squares are the measurements from N-body simulations of \citet{tin} and the solid line is the analytical fit to these predictions given by the above authors. The line of the analytical fit was derived using (51) and $\Delta =178$ in  Eqs (B1), (B2), (B3) and (B4) of the appendix of \citet{tin}, for the evaluation of constants A,a,b,c in Eq (51). Dashed-dot-dot line is the prediction for the SC model where ${\cal{F}}(S,z)$ is given by Eq.37.  Dots are the prediction of the \citet{shto02} approximation for the EC model (Eq. 38). Thin dashes are the prediction of the path integral approximation used in this work (Eq.28). In Eqs (28) and (38) we used for $(S_0,\delta_0)=(0,0)$ . Note that the Eqs (28), (38) and the numerical solution are almost indistinguishable from one another.Thick dashes show the numerical solution of (A5), for the ellipsoidal barrier, \citet{zhhu06}.}\label{fig1}
 \end{figure}
  \begin{equation}
  N(M,z)\mathrm{d}M=-\frac{\rho_b(z)}{M}{\cal{F}}(S,z)\frac{\mathrm{d}S(M)}{\mathrm{d}M}\mathrm{d}M
  \end{equation}
  that can be written in the form:
    \begin{equation}
  N(M,z)\mathrm{d}M=-\frac{\rho_b(z)}{M}\left[2S{\cal{F}}(S,z)\right]\frac{\mathrm{d}\ln(S^{-1/2})}{\mathrm{d}M}\mathrm{d}M
  \end{equation}
  We note here that some quantities in the above Eq. can be evaluated from the results of N-body simulations. \citet{shto99} showed that the combination $M\frac{N(M,z)}{\rho_b(z)}\frac{\mathrm{d}M}{\mathrm{d}\ln[\sigma^{-1}(M,z)]}$ has an almost universal behavior , that is independent of redshift and cosmology, a result that was confirmed by large numerical simulations  as those of \citet{tin}. We recall that $\sigma^2(M,z)$ is the variance at mass scale $M$ at redshift $z$. In the linear regime of the evolution it obeys the relation $\sigma(M,z)=\sigma(M,z=0)D(z)=1.686\sigma(M,z=0)B_{sc}(z)$.
  Introducing the variable $\nu\equiv B_{sc}(z)\sqrt{S}$, for constant $z$, we can write
  \begin{equation}
   \frac{\mathrm{d}S}{\mathrm{d}\nu}=2\frac{S}{\nu}
  \end{equation}
  Assuming a distribution function ${\cal{F}}_{\nu}$ of the variable $\nu$ and combining, for constant $z$, the fundamental law of probabilities,
  \begin{equation}
  {{\cal{F}}}(S,z)\mathrm{d}S=-{\cal{F}}_{\nu}(\nu)\mathrm{d}\nu
  \end{equation}
  with (45), we get:
  \begin{equation}
  \nu{\cal{F}}_{\nu}(\nu)=2S{{\cal{F}}}(S,z)
  \end{equation}
  and using (44) we have:
  \begin{equation}
  \nu{\cal{F}}_{\nu}(\nu)=\frac{M}{\rho_b(z)}N(M,z)\frac{\mathrm{d}M}{\mathrm{d}\ln(S^{-1/2})}
  \end{equation}
  \begin{figure}
 \includegraphics[width=18cm]{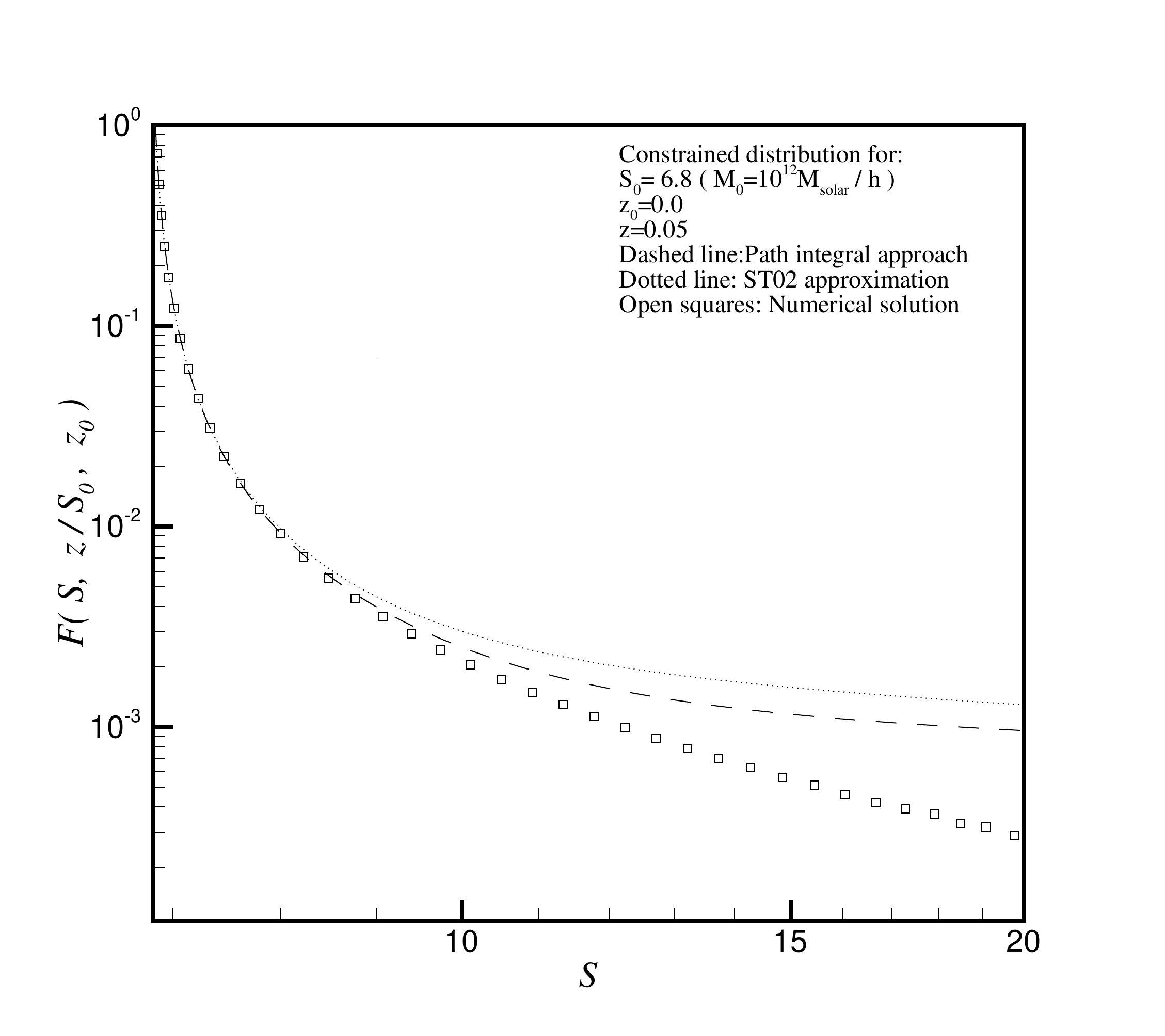}
 \caption{ Constrained distributions, ${\cal{F}}(S,z /S_0, z_0)$ for $S_0=6.8$ that corresponds to $M_0=10^{12}h^{-1}M_{\sun}$, $z=0.05$ and $z_0=0$. We used values of $S$ in the range $[S_0,S_{max}]$ where  $S_{max}=S(10^{-2}M_0) \simeq 22$.  All lines are predictions for the ellipsoidal barrier. Dotted line corresponds to the ST02 approximation, dashed line to the path approach and squares are from the numerical solution of (A5).}
\label{fig2}
 \end{figure}
         Taking into account that $B_{SC}(z)$ and $\sqrt{S}$ evolve with time in the same way according to the linear theory, the quantities  $\nu$ and $\nu{\cal{F}}_{\nu}(\nu)$ are time independent. $\nu{\cal{F}}_{\nu}(\nu)$ is called the ``multiplicity function".
    For SC model, using (37) it is easy to show that $2S{\cal{F}}_{SC}(S,z)=\frac{\sqrt{2}}{\pi}\nu\exp[-\frac{{\nu}^2}{2}]$ while for the EC model of \citet{shto02} we can write:
   \begin{equation}
   2S{\cal{F}}_{EC}(S,z)=\nu\frac{\sqrt{2}}{\sqrt{\pi}}|w_1+\frac{w_2[1+g(\gamma)]}{{\nu}^{2\gamma}}|
   \exp\left[-\frac{1}{2}{\nu}^2[w_1+\frac{w_2}{{\nu}^{2\gamma}}]^2\right]
   \end{equation}
   where
   \begin{equation}
  g(\gamma)=
  \mid 1-\gamma +\frac{\gamma (\gamma
  -1)}{2!}-...-\frac{\gamma(\gamma-1)\cdot \cdot \cdot
  (\gamma-4)}{5!} \mid
  \end{equation}
      Numerical experiments (see for example \citet{tin}) show that $\nu{\cal{F}}_{\nu}(\nu)$ is indeed nearly a constant. From their N-body simulations the above authors have shown that:
  \begin{equation}
   2S{\cal{F}}(S,z)_{n-body}=A\left[\left(\frac{\sqrt{S}}{b}\right)^{-a}+1 \right]e^{-\frac{c}{S}}
   \end{equation}
   where the constants $A, a, b$ and $c$ are given by the Eqs (B1), (B2), (B3) and (B4) of \citet{tin}.\\
    \begin{figure}
\includegraphics[width=18cm]{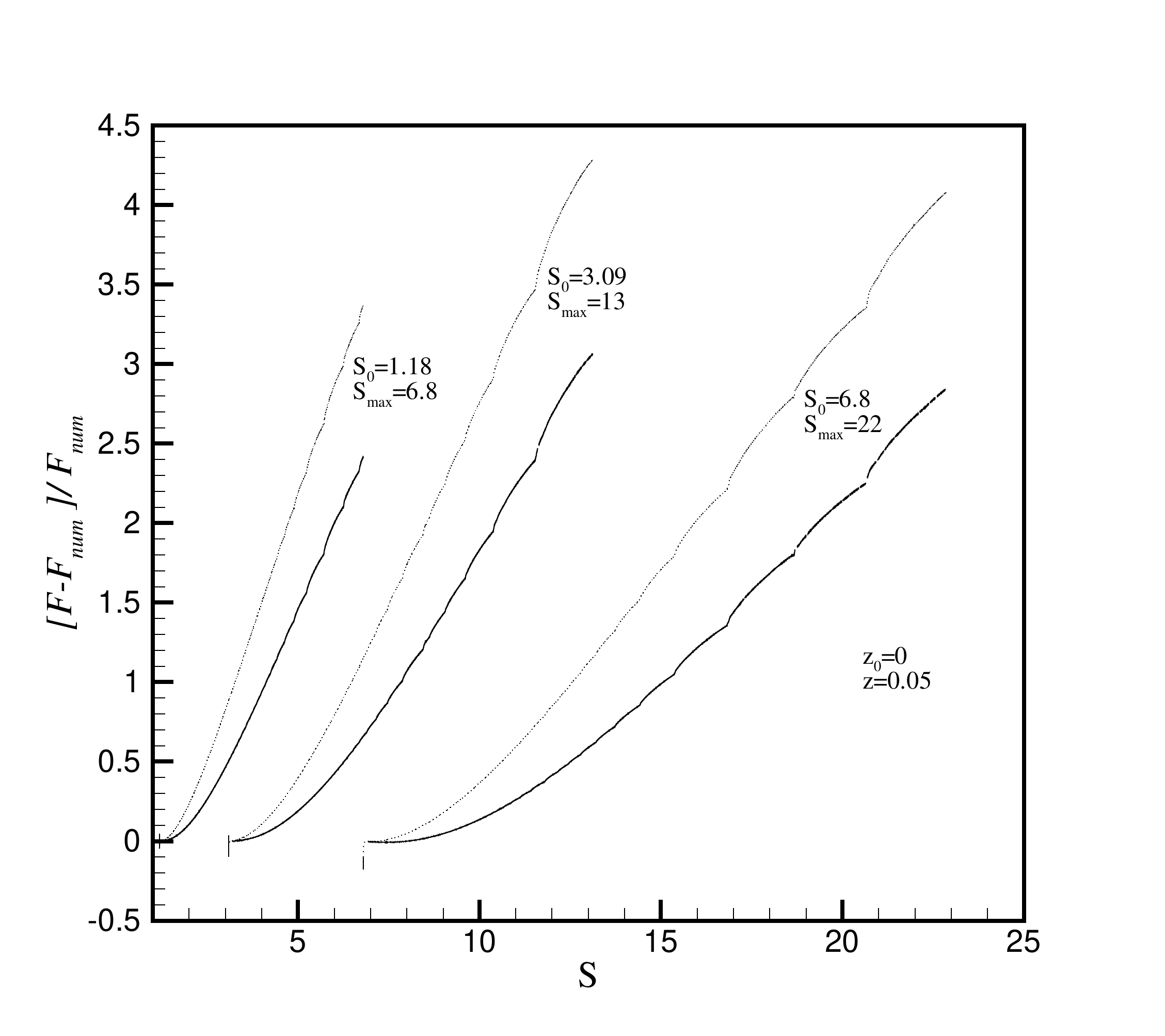}
\caption{The relative difference $ (\Delta {\cal{F}}/{\cal{F}})_{model}\equiv{(\cal{F}}_{model}-{\cal{F}}_{num})/({\cal{F}}_{num})$, where the index $model$ stands for the path or the ST02 approach and the index $num$ is for the numerical solution, as a function of $S$. Solid lines stand for the path approach while dotted lines for the ST02 approximation. All curves correspond to $z=0.05$ and $z_0=0$ but are predicted for different ranges of mass. The following values of $M$, $ 10^{10}M_{\sun}/h, 10^{11}M_{\sun}/h, 10^{12}M_{\sun}/h, 10^{13}M_{\sun}/h,10^{14}M_{\sun}/h$ correspond to $S=1.18, 3.09, 6.8, 13$ and $22$, respectively}\label{fig3}
 \end{figure}
       In what follows, we have assumed a flat model for the Universe with
  present day density parameters $\Omega_{m,0}=0.3$ and
   $ \Omega_{\Lambda,0}\equiv \Lambda/3H_0^2=0.7$ where
  $\Lambda$ is the cosmological constant and $H_0$ is the present day value of Hubble's
  constant. We have used the value $H_0=100~\mathrm{hKMs^{-1}Mpc^{-1}}$
  and a system of units with $m_{unit}=10^{12}M_{\odot}h^{-1}$,
  $r_{unit}=1h^{-1}\mathrm{Mpc}$ and a gravitational constant $ G=1$. At this system of units
  $H_0/H_{unit}=1.5276.$\\
 Regarding  the power spectrum, we  employed the $\Lambda CDM$ formula proposed by
  \citet{smet98}. The power spectrum is normalized for $\sigma_8\equiv\sigma(R=8h^{-1}\mathrm{Mpc})=0.9$.\\
\begin{figure}
\includegraphics[width=18cm]{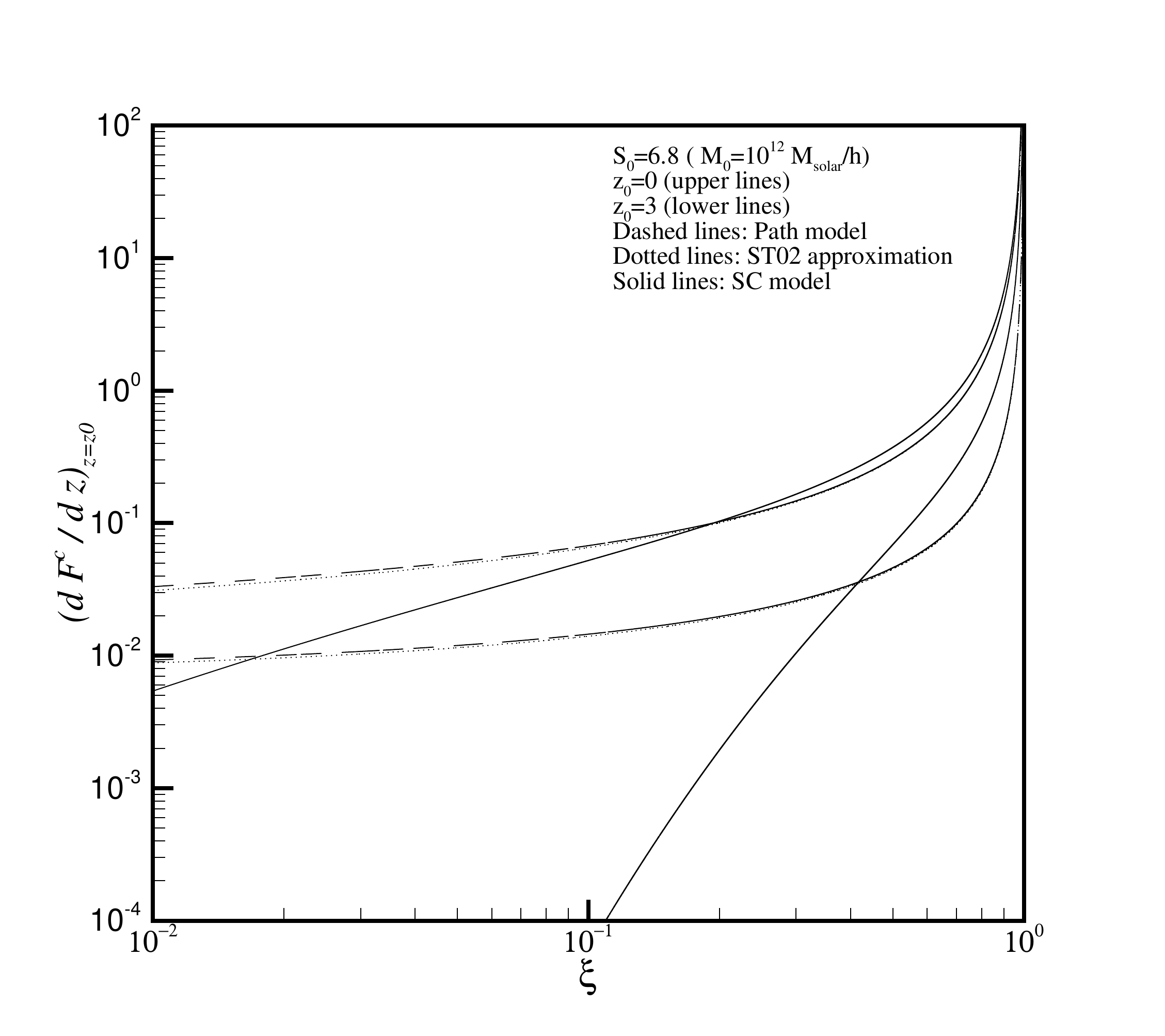}
\caption{ The values of $\frac{\mathrm{d}}{\mathrm{d}z}{{\cal{F}}_{SC}^c(S_0,z_0/S,z)}\mid_{z_0}$ as a function of $\xi$ (the ratio of the mass of the progenitor to the mass of the descendant halo), for $S_0=6.8$. This Fig. is for $z_0=0$ and $z_0=3$. Dashed lines correspond to the path approach model, dotted lines to the ST02 approximation while solid lines correspond to the SC model. Upper lines correspond to $z_0=0$ and lower lines to $z_0=3$.}\label{fig4}
 \end{figure}
   In Fig. 1  we have plotted multiplicity functions, $\nu{\cal{F}}_{\nu}(\nu)$. Squares are the predictions from N-body simulations of \citet{tin} and the solid line is the analytical fit to these predictions given by the above authors. Dashed-dot-dot line is the prediction for the SC model, where ${\cal{F}}(S,z)$ is given by Eq.37.  Dots are the prediction of the \citet{shto02} approximation for the EC model (Eq. 38). Thin dashes are the prediction of the path integral approximation used in this work (Eq.28). Thick dashes show the numerical solution of (A5) for the ellipsoidal barrier, \citet{zhhu06}.
  We note that  the results of \citet{shto02} approximation and the results of the path integral approach approximate the results of N-body simulations better than the results of SC model, especially for values of $\nu \leq 1$. This happens not only for  mass  functions (\citet{shto02})  but for other characteristics too as for example the formation times of dark matter haloes (eg. \citet{liet03}, \citet{hipo06}).\\
   In Fig. 2 we present constrained distributions for different halo mass and different redshifts. These distributions are given by (40), (41) and (28) for the SC, ST02 and the path approach models, respectively. Constrained distributions are essential when dealing with the formation of dark matter haloes. Important characteristic of dark matter haloes, as formation times, merger rates etc., depend on the form of the constrained    distributions. In this figure --for fixed values of $ S_0, z_0$ and $z$ ( for $z > z_0$)-- the quantity ${\cal{F}}( S, z / S_0, z_0)$ is plotted as a function of  $S$ (for $S > S_0$). This quantity is calculated from both analytical approximations and numerical solutions of (A5). It is shown that for values of $S$ close to $S_0$ -that correspond to masses $M$ close to $M_0$- analytical and numerical methods give the same results, but for large values of $S$ - that correspond to values of mass $M$  much smaller than mass $M_0$- the numerical solutions - assumed to be the exact solutions-  give values for ${\cal{F}}$ that are significantly smaller than those of analytical approximations. In order to give a more accurate comparison between the results we calculated the relative difference $ (\Delta {\cal{F}}/{\cal{F}})_{model}\equiv{(\cal{F}}_{model}-{\cal{F}}_{num})/({\cal{F}}_{num})$, where the index $model$ stands for the path or the ST02 approach and the index $num$ is for the numerical solution.  Both ${\cal{F}}_{num}$ and ${\cal{F}}_{model}$ are constrained distributions $(S,z/S_0,z_0)$. We present results in Fig.3. Solid lines stand for the path approach while dotted lines for the ST02 approximation. All curves correspond to $z=0.05$ and $z_0=0$ but are predicted for different ranges of mass. The following values of $M$, $ M=10^{10}M_{\sun}/h, M=10^{11}M_{\sun}/h, M=10^{12}M_{\sun}/h,M=10^{13}M_{\sun}/h,M=10^{14}M_{\sun}/h$ correspond to $S=1.18, 3.09, 6.8, 13$ and $22$ respectively. The results for other redshifts, in the range  range $[0.3]$ that we studied, are similar and show that the predictions  of the path approach are closer to the - exact- curve ( predicted by numerical solutions)  than the predictions of ST02 approximation.\\
   We note that the relative difference is an increasing function of $S$ starting from zero at $S=S_0$. Thus the quantity $F_x$ that is defined by the relation $F_x=\int_{S_0}^{S_x}{\cal{F}}(S,z/S_0,z_0)\mathrm{d}S$, gives  the fraction of walks that start from the point $(S_0, B(S_0, z_0))$ and pass from the point $(S, B(S,z))$ with $S$ in the range $[S_0, S_x]$. We define as $S_x$ the value of $S$ that satisfies $(\Delta {\cal{F}}/ {\cal{F}})_{model}(S_x)=x/100$ and thus $F_x$ equals to the fraction of walks that agree with the exact solution, better than $x $ percent. For all values of $x$ we have $S_{x,path} > S_{x, ST}$ and $F_{x,path} > F_{x,ST}$. We have also calculated $M_{x}\equiv M(S_x)$.In the following tables we give some characteristic results for $x=10$.\\
    \begin{table}[ht]
   \caption{Results for $z_0=0, z=0.05$} 
    \centering 
    \begin{tabular}{c c c c c c c} 
    \hline\hline 
     Mass  &  $S_{10,ST}$ & $ S_{10,path} $ & $F_{10,ST} $& $ F_{10,path}$ &$M_{10,ST}$ &$M_{10,path}$\\ [0.5ex] 
     \hline 
    $10^{12}h^{-1}M_{\sun}$ & 8.463 & 9.647 & 0.9854 & 0.9894 &$0.48\times 10^{12}h^{-1}M_{\sun}$ & $0.31\times 10^{12}h^{-1}M_{\sun}$\\ 
     $10^{13}h^{-1}M_{\sun}$ & 3.970 & 4.486 & 0.9623 & 0.9670 &$0.51\times 10^{13}h^{-1}M_{\sun}$ & $0.35\times 10^{13}h^{-1}M_{\sun}$\\
     $10^{14}h^{-1}M_{\sun}$ & 1.686 & 1.976 & 0.9498 & 0.9570& $0.44\times 10^{14}h^{-1}M_{\sun}$ & $0.31\times 10^{14}h^{-1}M_{\sun}$\\
    \hline 
    \end{tabular}
    \label{table:nonlin} 
     \end{table}
     \begin{table}[ht]
   \caption{Results for $z_0=3, z=3.05$} 
    \centering 
    \begin{tabular}{c c c c c c c} 
    \hline\hline 
     Mass  &  $S_{10,ST}$ & $ S_{10,path} $ & $F_{10,ST} $& $ F_{10,path}$ &$M_{10,ST}$ &$M_{10,path}$\\ [0.5ex] 
     \hline 
    $10^{12}h^{-1}M_{\sun}$ & 9.025 & 10.536 & 0.9691 & 0.9754 & $0.38\times 10^{12}h^{-1}M_{\sun}$ & $0.22\times 10^{12}h^{-1}M_{\sun}$\\ 
     $10^{13}h^{-1}M_{\sun}$ & 4.458 & 5.383 & 0.9532 & 0.9613 & $0.36\times 10^{13}h^{-1}M_{\sun}$ & $0.21\times 10^{13}h^{-1}M_{\sun}$\\
     $10^{14}h^{-1}M_{\sun}$ & 1.954 & 2.475 & 0.9392 & 0.9504 & $0.31\times 10^{14}h^{-1}M_{\sun}$ & $0.18\times 10^{14}h^{-1}M_{\sun}$\\
    \hline 
    \end{tabular}
    \label{table:nonlin} 
     \end{table}
     It is clear from the above two tables that path integral approach agrees to the exact results for a significant larger range of $S$ than ST approximation. $S_{10, path}$ can be by 26 percent larger than  $S_{10, ST}$ as for example it can be seen in the last row of Table 2.  Additionally, the results of path approach extended to larger masses. As it can be seen the interval of masses $[M, M_0] \equiv [M(S), M(S_0)] $, (note that $S>S_0$),  with an accuracy better than 10 percent becomes larger for the path approach since it extends to smaller values of mass. On the other hand, the same Tables show that the improvement in $F_{10}$ is not significant. This improvement seems to be an increasing function of $M_0$ and is less than $1.2$ percent for the range of parameters we studied. Similar results were found for other values of $x$ too.

         \section{ Merger rates}
         Using Bayes rule we write:
         \begin{equation}
         {\cal{F}}^c(S_0,z_0/S,z)=\frac{{\cal{F}}(S,z/S_0,z_0){\cal{F}}(S_0,z_0)}{{\cal{F}}(S,z)}
         \end{equation}
         Using (29), (41) or (42) we can write:
         \begin{equation}
         {\cal{F}}^c(S_0,z_0/S,z)=\frac{1}{\sqrt{2\pi}}\left[\frac{S}{S_0(S-S_0)}\right]^{3/2}
         e^{-\frac{[S_0B(S,z)-S\delta_0]^2}{2SS_0(S-S_0)}}\Phi_m(S,S_0,z,\delta_0)
         \end{equation}
         where
         \begin{equation}
         \Phi_m(S,S_0,z,\delta_0)\equiv\frac{G_{m}(S,S_0,z,\delta_0)G_{m}(S_0,0,z_0,0)}{G_{m}(S,0,z,0)}
         \end{equation}
         The index m states for the following models: SC model, ST-EC model and the path model, P-EC.
         Thus:
         \begin{equation}
         G_{SC}(S,S_0,z,\delta_0)=B_{SC}(z)-\delta_0= B_{SC}(z)-B_{SC}(z_0)
         \end{equation}
         and
         \begin{figure}
\includegraphics[width=20cm]{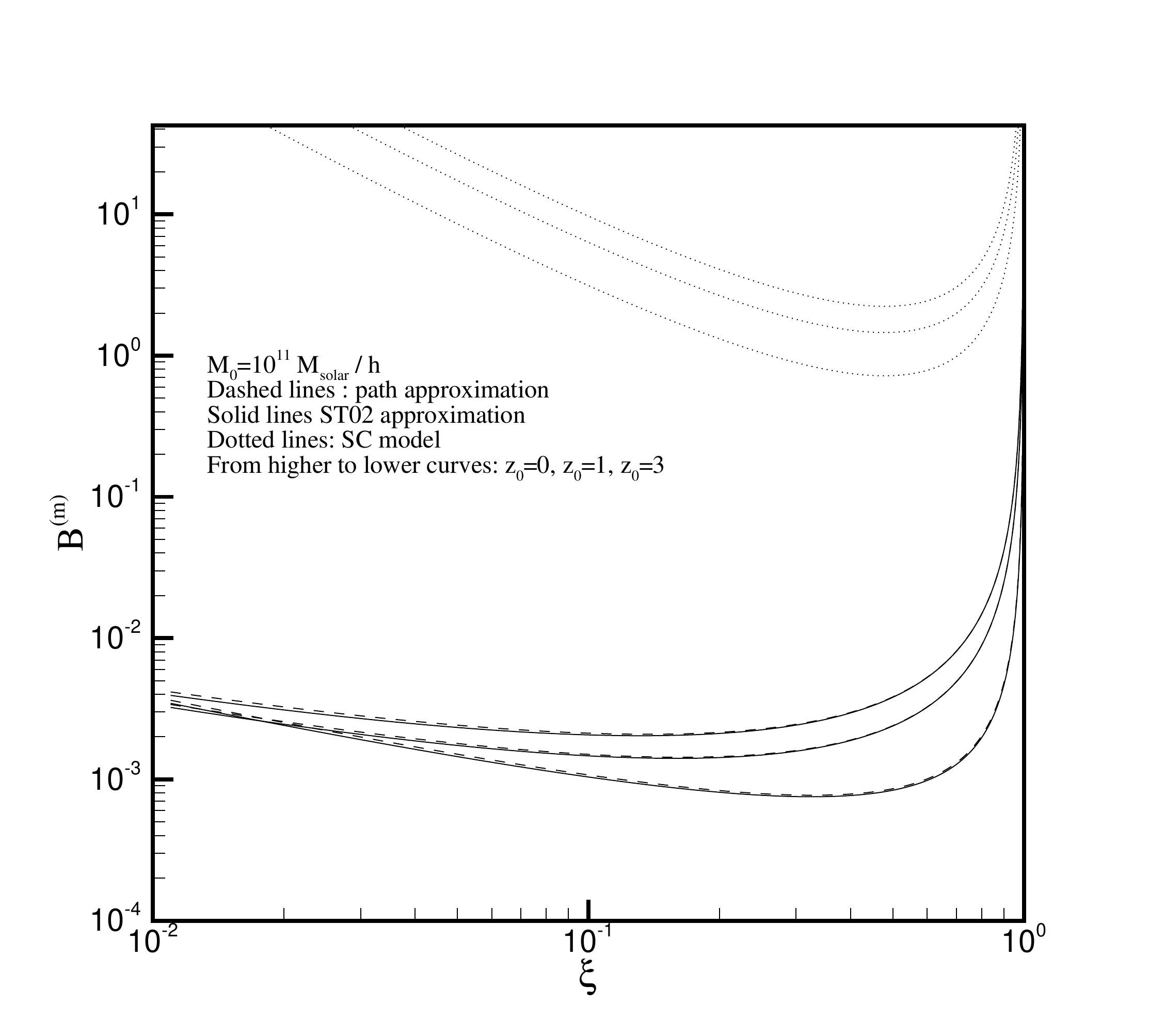}
\caption{Mean merger rates for $M_0=10^{11}M_{\sun}/h$ and for three different redshifts $z_0=0, z_0=1$ and $z_0=3$. Dashed lines (from top to bottom) correspond to the path approach for the above redshifts, respectively. Solid lines are the predictions of ST02 model and dotted lines are the predictions of SC model.}\label{fig5}
 \end{figure}
         \begin{equation}
         G_{ST-EC}(S,S_0,z,\delta_0)=\mid T(S,z/S_0,z_0)\mid
         \end{equation}
         while $G_{P-EC}(S,S_0,z,\delta_0)$ is given by (29). Differentiating (53) with respect to $z$  we have for the various models the following relations:\\
         For the SC model:
         \begin{eqnarray}
         \frac{\mathrm{d}}{\mathrm{d}z}{{\cal{F}}_{SC}^c(S_0,z_0/S,z)}=\frac{\delta_0}{\sqrt{2\pi}}
         \left[\frac{S}{S_0(S-S_0)}\right]^{3/2}
         \exp\left[-\frac{[S_0B_{SC}(z)-S\delta_0]^2}{2SS_0(S-S_0)}\right]\times\nonumber\\
         \left[\frac{\delta_0}{B_{SC}^2(z)}-\left(1-\frac{\delta_0}{B_{SC}(z)}\right)\frac{S_0B_{SC}(z)-S\delta_0}{S(S-S_0)}\right]
         \frac{\mathrm{d}B_{SC}(z)}{\mathrm{d}z}
         \end{eqnarray}
         For the $ST-EC$ model,\\
          \begin{eqnarray}
        \frac{\mathrm{d}{\cal{F}}_{ST-EC}^c(S_0,z_0/S,z)}{\mathrm{d}z}={\cal{F}}_{ST-EC}^c(S_0,z_0/S,z)\times\nonumber\\
        \left[\frac{S B(S_0,z_0)-S_0B(S,z)}{S(S-S_0)}\frac{\partial B(S,z)}{\partial z}+\frac{T(S,z)}{T(S,z/S_0,z_0)}\frac{\mathrm{d}}{\mathrm{d}z}\left(\frac{T(S,z/S_0,z_0)}{T(S,z)}\right)\right]
        \end{eqnarray}
        and for the $P-EC$ model:
         \begin{eqnarray}
        \frac{\mathrm{d}{\cal{F}}_{P-EC}^c(S_0,z_0/S,z)}{\mathrm{d}z}={\cal{F}}_{P-EC}^c(S_0,z_0/S,z)\times\nonumber\\
        \left[\frac{SB(S_0,z_0)-S_0B(S,z)}{S(S-S_0)}\frac{\partial B(S,z)}{\partial z}+\frac{G_{P-EC}(S,0,z,0)}{G_{P-EC}(S,S_0,z,\delta_0)}\frac{\mathrm{d}}{\mathrm{d}z}
        \left(\frac{G_{P-EC}(S,S_0,z,\delta_0)}{G_{P-EC}(S,0,z,0)}\right)\right]
        \end{eqnarray}
        In Fig 4 we present  $\frac{\mathrm{d}}{\mathrm{d}z}{{\cal{F}}_{SC}^c(S_0,z_0/S,z)}\mid_{z_0}$ as a function of $\xi$ for  $S_0=6.8$ ( that corresponds to $M_0=10^{12}M_{\Sun}/h$) and $z_0=0$, $z_0=3$.  Details about the particular values of the parameters are written in the figure caption. The results  show some interesting features:\\
         a) SC model gives completely different results. Only for large values of $\xi$, that means for large progenitors, the results of SC model are somehow close to those predicted by the models that use the ellipsoidal barrier. For small values of $\xi$, that means for small progenitors SC and EC models lead to completely different results.\\
         b) From the study of the same quantity for different values of the parameters mass $M_0$ and redshift $z_0$ we conclude that the  predictions of path approach and the ST02 approximation are close for any redshift and mass in the range $[0,3.]$ and $[10^{12}M_{\Sun}/h,10^{14}M_{\Sun}/h]$ respectively.\par
         We define as merger rate at redshift $z_0$ the quantity $R$,
         \begin{equation}
         R(M\rightarrow M_0/z_0)\mathrm{d}M_0\equiv\frac{\mathrm{d}f}{\mathrm{d}z}(M_0,z_0/M,z)\mid_{z=z_0}\mathrm{d}M_0
         =\frac{\mathrm{d}{\cal{F}}^c}{\mathrm{d}z}(S_0,z_0/S,z)\mid_{z=z_0}\mathrm{d}S_0
         \end{equation}
         $R(M\rightarrow M_0/z_0)$ equals to a fraction of the (total) mass that belongs to haloes of masses $M$. It is this fraction of mass that merges instantaneously to form at $z_0$ haloes of mass $M_0$. The quantity  $R(M\rightarrow M_0/z_0)f(M,z_0)\mathrm{d}M$ expresses the mass that belongs to haloes of mass $M,M+\mathrm{d}M$, and  merges instantaneously to form at $z_0$ haloes of mass $M_0$, as a fraction of the total mass of the Universe. Multiplying $R(M\rightarrow M_0/z_0)f(M,z_0)\mathrm{d}M$ by $V_{un}\rho_b/M$ and dividing by $V_{un}f(M_0,z_0)\rho_b/M_0$ where $V_{un}$ is the volume of the universe, we find:
         \begin{equation}
         \frac{N_m(M,z_0)}{N_m(M_0,z_0)\mathrm{d}z}=\frac{M_0}{M}\frac{f(M,z_0)}{f(M_0,z_0)}R(M\rightarrow M_0/z_0)\mathrm{d}M
         \end{equation}
                 In the above relation $N_m(M,z_0)$ is the number of haloes of mass in the range $M,M+\mathrm{d}M$ that merge at $z_0$ and form haloes of mass $M_0$ while $N_m(M_0,z_0)$ is the number of haloes of mass $M_0$ present at $z_0$. Using (33), we write:
         \begin{equation}
         \frac{N_m(M,z_0)}{N_m(M_0,z_0)\mathrm{d}z}=\frac{M_0}{M}\frac{{\cal F}(S,z_0)}{{\cal F}(S_0,z_0)}
         \frac{\mathrm{d}S}{\mathrm{d}M}\frac{\mathrm{d}}{\mathrm{d}z}{\cal F}^c(S_0,z_0/S,z)\mathrm{d}M
         \end{equation}
         In a merger procedure as above, the halo resulting from mergers between smaller ones, a halo with mass $M_0$ in our notation, is called a descendant halo. Haloes with smaller masses that merge to form $M_0$ are called progenitors. It is convenient for comparing merger rates for different descendant haloes to use, instead of the masses $M$ and $M_0$, the ratio $\xi\equiv M/M_0$ and thus (61) is written:
         \begin{equation}
         B^{(m)}_{M_0,z_0}(\xi)\equiv \frac{N_m(M,z_0)}{N_m(M_0,z_0)\mathrm{d}z\mathrm{d}\xi}=\frac{M^2_0}{M}\frac{{\cal{F}}(S,z_0)}{{\cal{F}}(S_0,z_0)}
         \left[\frac{\mathrm{d}{\cal{F}}^c(S_0,z_0/S,z)}{\mathrm{d}z}\right]_{z_0}
         \frac{\mathrm{d}S}{\mathrm{d}M}
         \end{equation}
                 and it  expresses  the mean merger rate. It measures the mean number of mergers per halo, per unit redshift, for descendant haloes of mass $M_0$ with mass ratio $\xi$ as defined above.\\
                 The mean merger rates for the three models are:
                  \begin{figure}
 \includegraphics[width=20cm]{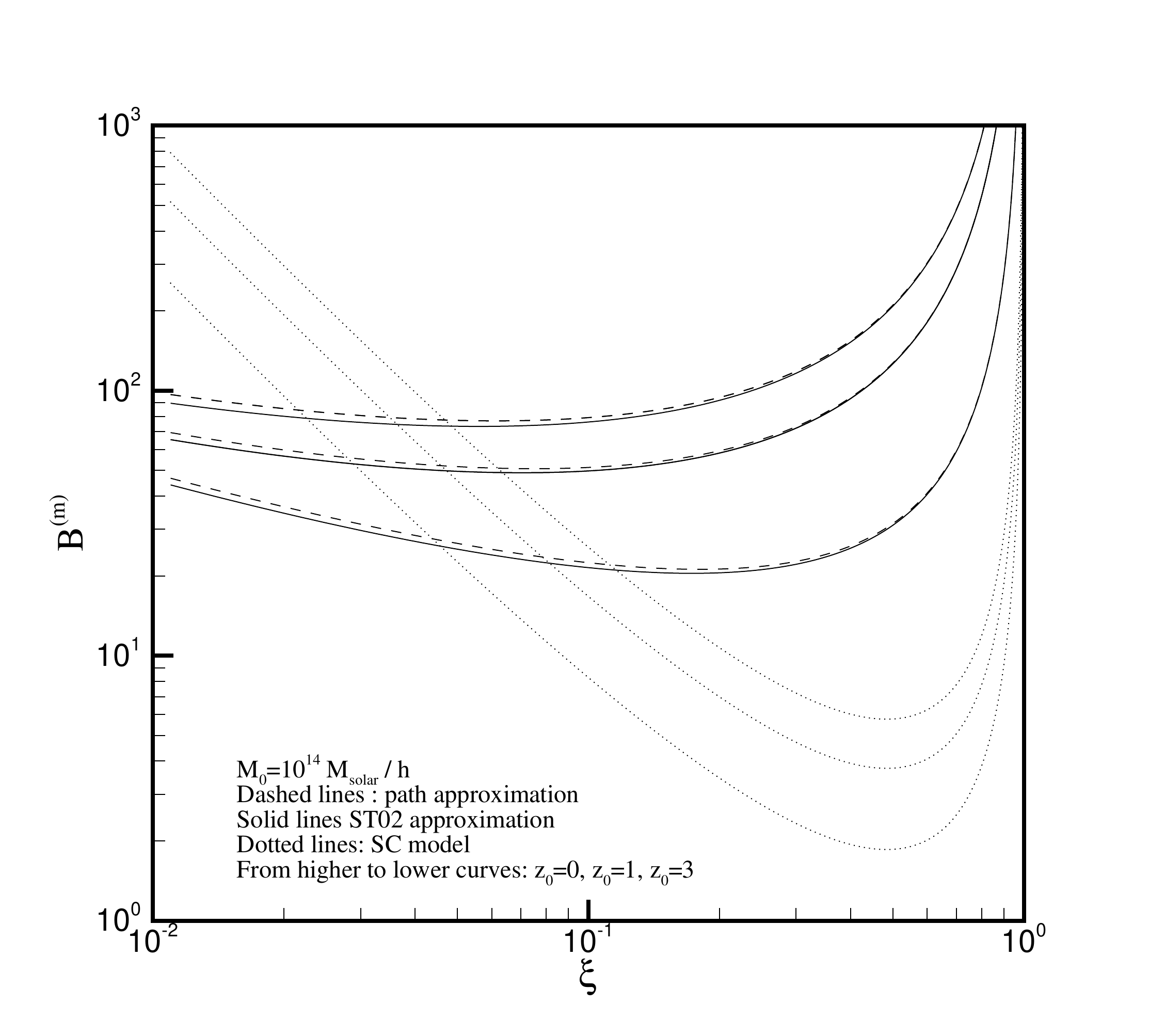}
\caption{As in Fig.5 but for $M_0=10^{14}M_{\sun}/h$.}\label{fig10}
 \end{figure}
         \begin{equation}
        B^{(m),SC}_{M_0,z_0}(\xi)=\frac{1}{\sqrt{2\pi}}\frac{M^2_0}{M}(S-S_0)^{-3/2}
        |\frac{\mathrm{d}S}{\mathrm{d}M}|\left[\frac{\mathrm{d}B_{SC}(z)}{\mathrm{d}z}\right]_{z_0}
        \end{equation}
        \begin{equation}
        B^{(m),ST-EC}_{M_0,z_0}(\xi)=\frac{M^2_0}{M}(S_0/S)^{3/2}
        |\frac{T(S,z_0)}{T(S_0,z_0)}|e^{\Delta B}\left[\frac{\mathrm{d}
       {\cal{F}}_{ST-EC}^c}{\mathrm{d}z}\right]_{z_0}\mid\frac{\mathrm{d}S}{\mathrm{d}M}\mid
        \end{equation}
         \begin{equation}
        B^{(m),P-EC}_{M_0,z_0}(\xi)=\frac{M^2_0}{M}\left(\frac{S_0}{S}\right)^{3/2}
       \frac{G_{P-EC}(S,0,z_0,0)}{G_{P-EC}(S_0,0,z_0,0)}e^{\Delta B}
       \left[\frac{\mathrm{d}
       {\cal{F}}_{P-EC}^c}{\mathrm{d}z}\right]_{z_0}\mid\frac{\mathrm{d}S}{\mathrm{d}M}\mid.
        \end{equation}
        where $\Delta B\equiv \frac{B^2_{EC}(S_0,z_0)}{2S_0}-\frac{B^2_{EC}(S,z_0)}{2S}.$\\
        In Figs 5 and 6 we present mean merger rates derived by Eqs 64, 65 and 66 for different redshifts and masses of descendant haloes. Details are written in figures captions. It is clear that for small haloes the predicted merger rates of SC model are far from those of the EC model that predicted by path approach (see Fig. 5). For increasing halo masses  the agreement becomes better. We note that analyzing the results of N-body simulations \citet{fak} and \citet{stew} found analytical approximations for merger rates of dark matter haloes but their definition of $\xi$ differs to the one  used above. In their papers, $\xi$ is not defined as $\xi \equiv M/M_0$ but as $\xi \equiv M/M_{MMP}$, where $M_{MMP}$ is the most massive progenitor during the merge process. With this definition of $\xi$ merger rates cannot be found analytically. Their derivation requires the construction of merger-trees, that is a complex process with its own difficulties, (e.g. \citet{nede08}). In \citet{hiot11}, using merger-tress, it  was shown that SC model approximates better merger rates for small haloes while merger rates of larger haloes are approximated better from models that use the ellipsoidal barrier  (see Fig.9 therein).\\
        A direct detailed comparison of mean merger rates derived by Eqs 64, 65 and 66 -where  $\xi \equiv M/M_0$- with those predicted from N-body simulations is the next step in our research.
\section{Conclusions and Discussion}

         The power tool of path integrals is used in order to approximate the problem of the formation of dark matter haloes. The initial density field is assumed to be Gaussian and it is smoothed by a filter that is sharp in \textbf{k} space. Using these assumptions, the results show that decreasing the radius of the filter, the smoothed density evolves according to a Langevin equation of a form that describes a Markovian stochastic process. Thus, it was able to calculate first crossing distributions for two -well known in the literature- namely a flat barrier corresponding to the Spherical Collapse model and a moving barrier corresponding to the Ellipsoidal Collapse Model. Mass functions of dark matter haloes, predicted by the above first crossing distributions, are compared with numerical solutions -that are considered as exact- and with the predictions of N-body simulations. It is shown that for the EC model -that is more promising than the SC model-  the prediction of path approach is close to the results of \citet{shto02}.\\
          Although in the calculation of  ${\cal{F}}$ we used only two terms from an expansion (see Eqs. 28 and 29), the resulting constrained first crossing distributions from the path approach are closer to the ones predicted by the numerical solutions than the predictions of \citet{shto02} are. We note that the infinite sum in (29) is accurately approximated using up to 12 terms. This shows, that the analytical formula given in Eq. 28, that resulted from the analytical procedure described in the text, improves the empirical formula (see Eqs. 41 and 42) of \citet{shto02}. We note that the predictions of the analytical formula of ST02 are predicted using about 6-8 terms of the sum. \\
         Merger rates have also been calculated. Merger rates resulting from the path integral  approach are close to those predicted by \citet{shto02} approximation and no significant improvement appears. The use of additional terms in Eqs. 28 and 29 -that could bring the results of path approach closer to that of the numerical solutions of the integral equation (A5)- as well as a detailed comparison with the merger rates resulting from N-body simulations are clearly required. Both actions are in progress.
\section{Acknowledgements}
\ We acknowledge  J.Tinker for kindly providing the results of their N-body simulations. We also thank  Dr. G. Kospentaris  and K. Konte for assistance in manuscript preparation and the \textit{Empirikion} Foundation for financial support.

\newpage
\appendix

    \section{Appendix A}
     We denote by $P(S,\delta/S_0,\delta_0)\mathrm{d}\delta$ the probability a trajectory that starts from the point $(S_0,\delta_0)$ passes at ``time" $S$ between $\delta,\delta+\mathrm{d}\delta$ without crossing the barrier $B(S,z)$ between $S_0$ and $S$. On the other hand  ${\cal{F}}(S,z/S_0,\delta_0)\mathrm{d}S\equiv{\cal{F}}(S,B(S,z)/S_0,\delta_0)\mathrm{d}S$ equals to the probability a trajectory that passes from the point $(S_0,\delta_0)$ crosses the barrier of height $B(S,z) > \delta_0$, for the first time, between $S,S+\mathrm{d}S$. Consequently, $P_1$ given by
     \begin{equation}
     P_1=\int_{S_0}^S{\cal{F}}(S',B(S',z) /S_0,\delta_0)\mathrm{d}S'
     \end{equation}
     is the probability the trajectory has crossed the barrier before $S$ while $P_2$,
     \begin{equation}
     P_2=\int_{-\infty}^{B(S,z)}P(S,\delta/S_0,\delta_0)\mathrm{d}\delta
     \end{equation}
     is the probability the trajectory was always at values smaller than $B(S,z)$ and thus it has not crossed the barrier for ``time" $< S$.
     Obviously $P_1+P_2=1$. We also assume that the transition probability $P_0$, in the absence of any barrier, is a normal Gaussian given by:
     \begin{equation}
     P_0(S,\delta/S_0,\delta_0)=\frac{1}{\sqrt{2\pi(S-S_0)^3}}\exp\left[-\frac{(\delta-\delta_0)^2}{2(S-S_0)}\right]
     \end{equation}
     The presence of the barrier amplifies $P$ in the following way:
     \begin{equation}
      P(S,\delta/S_0,\delta_0)= P_0(S,\delta/S_0,\delta_0)-
      \int_{S_0}^{S}{\cal{F}}[S',B(S',z)/S_0,\delta_0] P_0[S,\delta/S',B(S',z)]\mathrm{d}S'
      \end{equation}
       Using (A3) and (A4) it can be proved \citep{zhhu06}, that for an arbitrary barrier,  ${\cal{F}}$ satisfies the following integral equation:
   \begin{equation}
   {\cal{F}}[S,B(S,z)/S_0,\delta_0]=g_1(S,\delta_0,S_0)+\int_{S_0}^Sg_2(S,S'){\cal{F}}[S',B(S',z)/\delta_0,S_0]\mathrm{d}S'
   \end{equation}
   where:
   \begin{equation}
   g_1(S,\delta_0,S_0)=\left[\frac{B(S,z)-\delta_0}{S-S_0}-2\frac{\partial
   B(S,z)}{\partial S}\right]P_0[S,B(S,z)/S_0,\delta_0]
   \end{equation}
   \begin{equation}
   g_2(S,S')=\left[2\frac{\partial B(S,z)}{\partial S}-\frac{B(S,z)-B(S',z)}{S-S'}\right]P_0[S,B(S,z)/ S'B(S',z)]
   \end{equation}
     In the case of a linear barrier Eq.(A5) admits an analytic solution. If $B(S,z)=p_c(z)+q_c(z)S$, where  $p_c$ and $q_c$ are functions of the  redshift $z$, the solution is written:
   \begin{equation}
   {\cal{F}}[S, z / S_0,\delta_0]=\frac{B(S_0,z)-\delta_0}{\sqrt{2\pi(S-S_0)^3}
   }\exp\left[-\frac{[B(S,z)- \delta_0]^2}{2(S-S_0)}\right]
   \end{equation}
   Thus, the spherical model which is of the form $B(S,z)=B(z)=1.686/D(z)$ leads to the solution:
   \begin{equation}
   {\cal{F}}_{SC}(S,z/S_0,z_0)=\frac{B(z)-\delta_0}{\sqrt{2\pi(S-S_0)^3}}\exp\left[-\frac{[B(z)-\delta_0]^2}
   {2(S-S_0) }\right]
   \end{equation}
   Eq. (A5) admits a simple numerical solution. First, we use a grid of points  $S_k=S_0+k\Delta S$ for  $k=0,1...N$ and $\Delta S=(S-S_0)/ N$ (thus $S_N=S$), for the
   interval $[S_0, S]$. Then, using the trapezoidal rule, we write (A5) as:
   \begin{equation}
   {\cal{F}}_i=g_1(S_i, \delta_0, S_0)+
   \frac{\Delta S}{2}\sum_{j=1}^{j=i}g_2(S_i,S_j-\frac{\Delta S}{2})[ {\cal{F}}_j+ {\cal{F}}_{j-1}]
   \end{equation}
   where we set ${\cal{F}}_k\equiv {\cal{F}}(S_k,z/S_0,z_0)$. Solving for ${\cal{F}}_i$ we have the solution of the integral equation:
   \begin{equation}
   {\cal{F}}_i=\frac{g_1(S_i, \delta_0, S_0)+ \frac{\Delta S}{2}\sum_{j=1}^{j=i-1}g_2(S_i,S_j-\frac{\Delta S}{2})[ {\cal{F}}_j+ {\cal{F}}_{j-1}]}{1-\frac{\Delta S}{2}g_2(S_i,S_i-\frac{\Delta S}{2})}
   \end{equation}
   The above Eq. holds for $i \geq 2$ while for $i=0$ or $i=1$ we have
   ${\cal{F}}_0=0$ and ${\cal{F}}_1=g_1(S_1, \delta_0, S_0)/[1-\frac{\Delta S}{2}g_2(S_1,S_1-\frac{\Delta S}{2})]$, respectively.\\
   We also used an iterative method for the solution of (A5). We denote by  ${\cal{F}}^{(l)}_i$ the $l^{th}$ approximation for the value ${\cal{F}}_i$. We use the initial conditions ${\cal{F}}^{(0)}_i=g_1(S_i, \delta_0, S_0)$ and the iterative method:
    \begin{equation}
   {\cal{F}}^{(l+1)}_i=g_1(S_i, \delta_0, S_0)+
   \frac{\Delta S}{2}\sum_{j=1}^{j=i}g_2(S_i,S_j-\frac{\Delta S}{2})[ {\cal{F}}^{(l)}_j+ {\cal{F}}^{(l)}_{j-1}]
   \end{equation}
   The iterations stop when for all values of $i=1,2..N$, the condition:
   \begin{equation}
    \mid({\cal{F}}^{(l+1)}_i-{\cal{F}}^{(l)}_i)/{\cal{F}}^{(l)}_i\mid < \varepsilon
    \end{equation}
    We used $\varepsilon = 10^{-5} $.  We note that the numerical solution of (A5) resulting from any of the above two methods is very sensitive to the number of grid points N. We used a large number of grid points N, $N=10^5$, to divide the interval $[S_0, S_{max}]$ where $S_0=S(M_0)$ and $S_{max}=S(10^{-2}M_0)$ so as the two numerical methods give identical results. The accuracy of these numerical results was checked by comparing them with the exact solutions that are available from the SC model. The agreement was found almost exact.
    \end{document}